\documentstyle[pre,aps,epsf,graphicx]{revtex}
\begin{document}

\draft

\title{Grand Canonical Model Predictions For Nuclear Fragmentation}
 
\author{C. B. Das$^{1,2}$, S. Das Gupta${^1,}\thanks{Corresponding author}$ and B. K. Jennings$^3$}

\address{$^1$Physics Department, McGill University, 
Montr{\'e}al, Canada H3A 2T8}
\address{$^2$ Variable Energy Cyclotron Centre, 1/AF Bidhannagar,
Kolkata 700 064, India}
\address{$^3$TRIUMF, 4004 Wesbrook Mall, Vancouver, Canada V6T 2A3}

\date{\today}

\maketitle

\begin{abstract}
The grand canonical ensemble has been used to make predictions for
composite yields using simple models for nuclear fragmentation.  
While this gives correct model prediction for high energy collisions,
it can give very inaccurate results at intermediate energy.

\end{abstract}


A very simple but very popular model for nuclear multifragmentation
is this: the nucleus is heated up and breaks up into many pieces
(composites and new produced particles if the energy is sufficient)
strictly according to phase space.  This occurs in
an expanded volume, about three or four times the normal volume.
Population strictly according to phase-space implies chemical
and thermal equilibrium.  For simplicity, we will omit new particle
production.  In such cases the number of dissociating particles
is fixed.  Nonetheless in the past it has been customary for
calculational simplicity to use the grand canonical ensemble
to describe multifragmentation 
\cite{Mekjian1,Gosset,Dasgupta1}.  In such a model (GCM), the number of
particles in the dissociating system is not constant, however one
can arrange to have the average number correspond to the actual
system.  If one is in the classical regime
(Fermi or Bose statstics degenerates into Maxwell-Boltzmann limit)
then the average yield of a composite in the ground state is given by
\begin{eqnarray}
\langle n_{i,j}(ground)\rangle =e^{i\beta\mu_z+j\beta\mu_n}f_{i,j}
\end{eqnarray}
where $i$ is the proton number, $j$ is the neutron number of the
composite, $\mu_z$ is the proton chemical potential, $\mu_n$ is
the neutron chemical potential
and  $f_{i,j}$ is given by:
\begin{eqnarray}
f_{i,j}=g\frac{V}{h^3}(2\pi mT)^{3/2}a^{3/2}\exp(\beta E_{i,j}) . \nonumber
\end{eqnarray}
Here $V$ is the volume within which the particle moves, $a=i+j$ is
the mass number of the composite, $m$ is the proton mass, 
$g$ is the spin degeneracy,
$E_{i,j}$ is the binding energy of the composite and the 
Maxwell-Boltzmann distribution of the momentum of the particle has
been integrated over.  Usually populations into any states of the composite,
ground and excited are included (a popular method of including the
excited states is to use the Fermi-gas approximation)
in which case $f_{i,j}$ is
replaced by $\omega_{i,j}$, the one particle partition function of
the particle.  Thus we have 
\begin{eqnarray}
\langle n_{i,j} \rangle =e^{i\beta\mu_z+j\beta\mu_n}\omega_{i,j}
\end{eqnarray}
The Wigner-Seitz approximation of the coulomb energy is usually included
\cite{Bondorf} and this can be incorporated in the $\omega_{i,j}$ by
replacing the coulomb self-energy $\frac{3i^2e^2}{5a}$ of the composite
$a(=i+j)$ to $\frac{3i^2e^2}{5a}(1-(\rho/\rho_0)^{1/3})$.
The chemical potentials $\mu$ are fixed from
\begin{eqnarray*}
\sum_{i,j}i\langle n_{i,j} \rangle &=& Z \\ \nonumber
\sum_{i,j}j\langle n_{i,j} \rangle &=& N \nonumber
\end{eqnarray*}
where $Z,N$ are the charge and neutron number of the dissociating system.
The connection between the GCM and the Canonical Model (CM), as described
in a textbook, is simple mathematics.  Let us denote the CM partition
function as $Q$ and the GCM partition function as $\tilde Z$, then
\begin{eqnarray*}
Q_{n_{i,j}} &=& \frac{(\omega_{i,j})^{n_{i,j}}}{n_{i,j}!} \ {\rm and} \\ \nonumber
\tilde Z &=& \prod_{i,j} \left [\sum_{n_{i,j}=0}^{\infty} e^{(i\beta\mu_z+j\beta\mu_n)n_{i,j}} Q_{n_{i,j}}\right] \\ \nonumber 
&=& \prod_{i,j}\exp \left [e^{(i\beta\mu_z+j\beta\mu_n)}\omega_{i,j} \right] . \nonumber
\end{eqnarray*}

\vskip -0.2 in
\begin{figure}
\epsfxsize=4.5in
\epsfysize=6.0in
\centerline{\epsffile{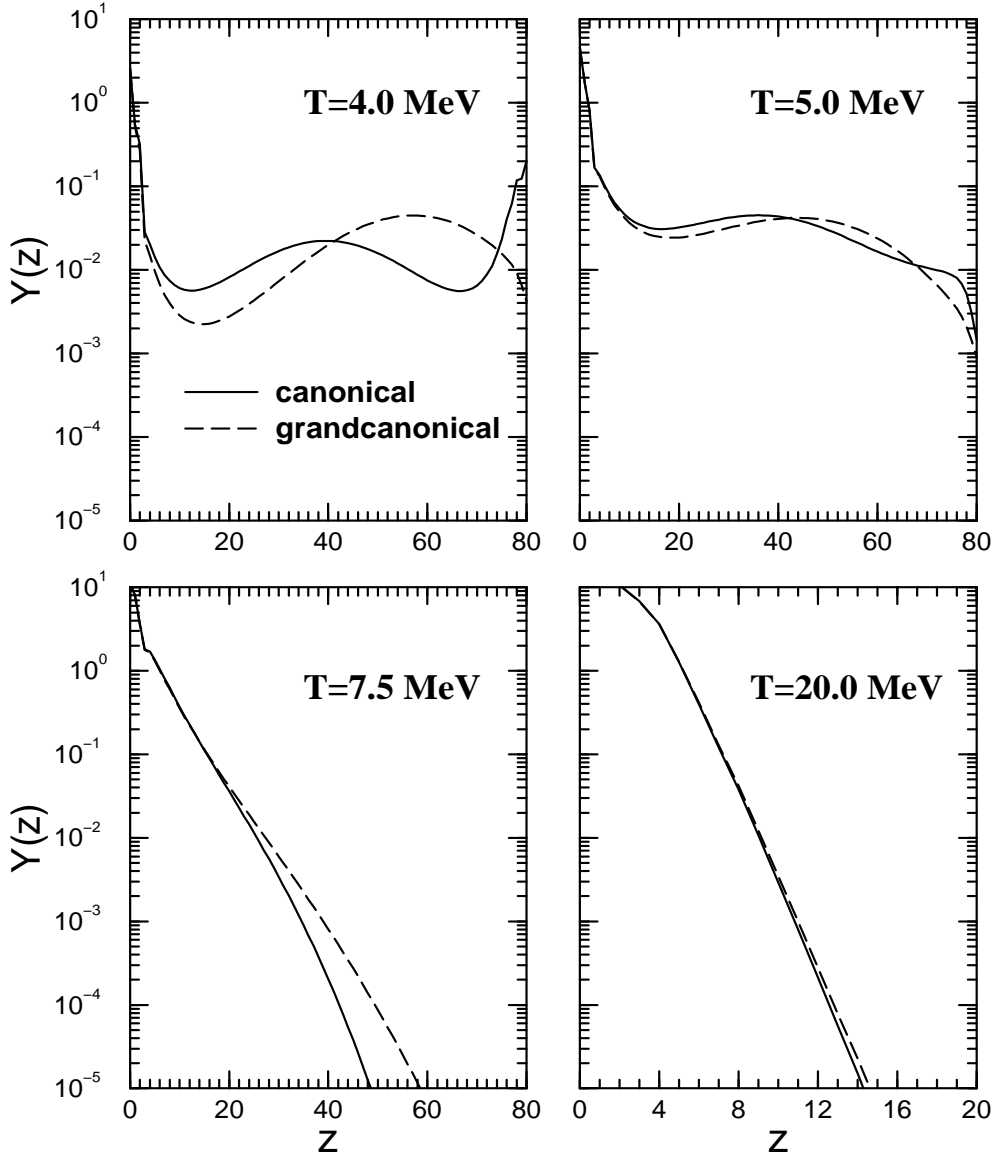}}
\vskip 0.8 true cm
\caption{\it The yields in the multifragmentation model using the Grand Canonical
ensemble and the canonical ensemble for $A=200, Z=80$.  Note that the
two ensembles give very different values at $T$=4 MeV.}
\end{figure}

Recently \cite{Dasgupta2,Bhat1}
it has become possible to use the canonical model
to calculate yields of fragmentation, whereas, in the past, the GCM
was universally used.  In the general case many composites are present
and hence

\begin{eqnarray}
Q_{Z,N}=\sum\prod_{i,j}\frac{\omega_{i,j}^{n_{i,j}}}{n_{i,j}!}
\end{eqnarray}
The sum is over all partitions of $Z, N$ into clusters and nucleons
subject to two constraints: $\sum_{i,j}in_{i,j}=Z$ and
$\sum_{i,j}jn_{i,j}=N$.  These constraints would appear to make
the computation of $Q_{Z,N}$ prohibitively difficult 
which used to be the primary reason for the use of the grand
canonical ensemble  where with two constants $\mu_z,\mu_n$ one
merely arranged the average values to be $Z$ and $N$.  It has
been recently realised that a recursion
relation exists which allows the computation of $Q_{Z,N}$ quite easy
on the computer even for large $Z$ or $N$ \cite{C}.  Three equivalent recursion
relations exist, any one of which could be used.  For example, one
such relation is 
\begin{eqnarray}
Q_{z,n}=\frac{1}{z}\sum_{i,j}i\omega_{i,j}Q_{z-i,n-j}
\end{eqnarray}

The average number of particles of the species $i,j$ is given by
\begin{eqnarray}
\langle n_{i,j} \rangle=\omega_{i,j}\frac{Q_{Z-i,N-j}}{Q_{Z,N}}
\end{eqnarray}
All nuclear properties are contained in $\omega_{i,j}$.  

\vskip -0.2 in
\begin{figure}
\epsfxsize=4.5in
\epsfysize=6.0in
\centerline{\epsffile{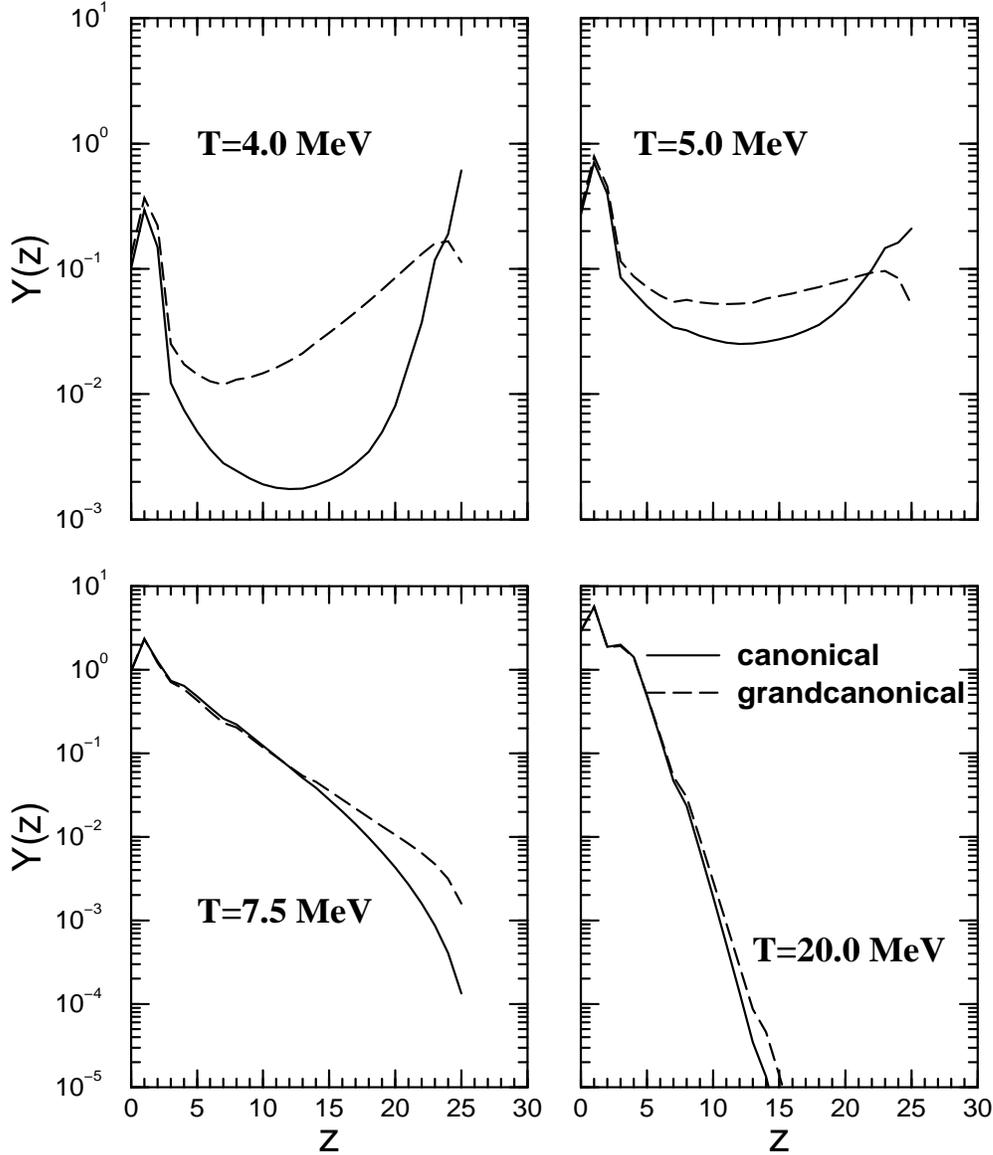}}
\vskip 0.8 true cm
\caption{\it The same as in Fig.1 except for the system $A=50, Z=25$.
Again note the discrepancies at $T$=4.0 MeV.}
\end{figure}

We are in a position to check, in the nuclear case, the grand canonical
predictions for yields with the canonical values where particle number 
conservation is strictly enforced.  We show this 
in Fig.1 for $A=200, Z=80, N=120$ (on the nuclear scale, a large system)
and in Fig.2 for $A=50, Z=25$, N=25.  
As there are too many composites,
we compare isotope yields (yields of the same $Z$ are added up and
then compared).  The GCM and CM predictions are
quite close for high temperatures ($T\ge $20 MeV) but at low 
temperatures ($\approx$ 4 MeV) in spite of 200 being a large number,
the GCM predictions are significantly
different.  One encounters such temperatures in intermediate energy
heavy ion collisions thus one would conclude that one should not use the GCM
in intermediate energies.  GCM has been used at 100 MeV/nucleon beam energy
in the lab \cite{Randrup} where it may be just adequate but it has also
been used at much lower energy \cite{Pal,Dass} where its usage is
very questionable.

One has to ask if, when the two predictions differ, are both of them wrong
or only one of them?  If we consider thermal and chemical equilibrium
to be the fundamental ingredient of the model, then the 
only way the present calculation in CM
can be wrong is, if the approximation to quantum statistics (eq.3) 
is invalid around
5 MeV.  Following a recent paper 
\cite{Jennings} we show this is not true and thus CM
model results continue to be good.  Here is a gist of the argument
from that paper.

We use large volumes (3 or 4 times the normal volume).  At low temperatures
($\approx$ 4 MeV) where one might imagine the approximation to fail, it
survives because different composites appear thus there is not enough
of any particular species to make (anti)symmetrisation an important
issue.  At much higher temperature, the number of protons and neutrons
increase but as is well-known, the $n!$ correction takes the approximate
partition function towards the proper one.  In a different world, the problem
could get very difficult.  Such a scenario would arise if the physics
was such that at low temperatures we only had neutrons and protons and
no composites.  An even worse situation would arise if we had only
neutrons (or protons).  With these preliminaries, let us proceed to
to estimate quantitatively the errors involved in actual cases that
one might encounter.  For simplicity only, in this section we will
not put in excited states of composites and we have not put in the Wigner-Seitz
correction although that could have been retained without much extra work.

A recursive relation similar to eq.4 exists even with incorporation
of quantum statistics but $\omega_{i,j}$'s are no longer one-particle
partition functions.  We illustrate this first with the example of only
protons filling up orbitals $i,j,k...$ in a box.  Now 
\begin{eqnarray}
lnQ_{gr}(\beta\mu) &=& \sum_iln(1+e^{\beta\mu-\beta\epsilon_i}) \nonumber \\
                    &=& \sum_i\sum_j\frac{(-)^{j-1}}{j}e^{j(\beta\mu-\beta\epsilon_i)}
\end{eqnarray}
The coefficient of $e^{\beta\mu k}$ is 
$x_k=\frac{(-)^{k-1}}{k}\sum_ie^{-k\beta\epsilon_i}$.  The canonical
partition function for $Z$ protons is very similar to eq.4:
\begin{eqnarray}
Q_Z(\beta)=\frac{1}{Z}\sum_{k=1}^Zkx_kQ_{Z-k}
\end{eqnarray}
Here $Q_0$ is 1.
When the expressions for $x_k$ are used in the above equation, 
orbitals are given occupancies greater
than one and then eliminated by subtraction.  This can lead to severe
round-off errors when applied to degenerate Fermi systems but will
not affect the applications here.  The number of protons $Z$ is given
by
\begin{eqnarray}
Z=x_1\frac{Q_{Z-1}}{Q_Z}+2x_2\frac{Q_{Z-2}}{Q_Z}+.....Zx_Z\frac{Q_0}{Q_Z}
\end{eqnarray}
For generalisation, we will call $x_k$ above as $y_{1,0}^{[k]}$.  The 
symbol 1,0 means it is a composite with 1 proton and 0 neutron.
The symbol $k$ means it is obtained from the $k-th$ term in the 
expansion; $y_{1,0}^{[k]}$ will contribute to $x_{k,0}$.

If instead we had a boson, a deuteron, for example, we would have
\begin{eqnarray}
ln[Q_{gr.can}(\beta,\mu_p,\mu_n)] & = & \sum_i-ln(1-e^{\beta\mu_p+\beta\mu_n}
e^{-\beta\epsilon_i}) \\  
& = &\sum_i\sum_j\frac{1}{j}e^{j(\beta\mu_p+\beta\mu_n-\beta\epsilon_i)}
\end{eqnarray}
Thus in the case of deuterons $y_{1,1}^{[k]}$ (which would contribute to
$x_{k,k}$) is given by $\sum_i\frac{1}{k}e^{-k\beta\epsilon_i}$. 
 
We can treat an assembly of protons, neutrons, deuterons, tritons...etc.
The recursive relation if the dissociating system has $Z$ protons
and $N$ neutrons is
\begin{eqnarray}
Q_{Z,N}=\frac{1}{Z}\sum_{i=1,Z,j=0,N}ix_{i,j}Q_{Z-i,N-j}
\end{eqnarray}
The average number of a composite with $i_1$ protons and $i_2$ neutrons
is given by
\begin{eqnarray}
<n_{i_1,i_2}>=y_{i_1,i_2}^{[1]}Q_{Z-i_1,N-i_2}/Q_{Z,N}+2y_{i_1,i_2}^{[2]}
Q_{Z-2i_1,N-2i_2}/Q_{Z,N}+...
\end{eqnarray}
Unless one is in an extreme degenerate fermi system, one can evaluate
the $y$ factors by replacing sums with integration.  For example,
$y_{1,0}^{[n]}=
\frac{(-)^{n-1}}{n}\sum_i e^{-n\beta\epsilon_i}$ where the sum is
replaced by
$\int e^{-n\beta\epsilon} g(\epsilon)d\epsilon
=2\frac{V}{h^3}(\frac{2\pi m}{n\beta})^{3/2}$.  Here 
$V$ is the available volume.  We have included the
proton spin degeneracy; $m$ is the proton mass.  
For the deuteron, $y_{1,1}^{[k]}= 
\frac{1}{k}\int e^{-k\beta\epsilon}g(\epsilon)d\epsilon$.  This is
$3\times 2^{3/2}\frac{V}{h^3}(\frac{2\pi m}{\beta})^{3/2}\frac{e^{k\beta E_b}}
{k^{5/2}}$ where $E_b$ is the binding energy of the deuteron.  It is clear
how to compute contributions from other composites.

We test the accuracy of the yields as calculated throughout the main
text by comparing with a calculation where the complete theory of
symmetrisation and antisymmetrisation is used.  Subject only to the
approximation that summation over discrete states has been replaced by
an integration over a density of states, the calculation is exact.
The results are taken from \cite{Jennings}.  We take the dissociating system
to have $Z$=25 and $N$=25.  The lowest temperature considered is 3
MeV (one might argue that at lower temperature a model of sequential
decay is more appropriate).  The highest temperature shown is 30 MeV.
We take a freeze-out volume in which the composites can move freely
in three times the volume of a normal nucleus with 50 nucleons.
Aside from neutrons and protons we allow the possibility of composites.
Spins and binding energies for deuteron, triton,
 $^3$He and $^4$He are taken from experiments.  For higher mass composites
the binding energy is taken from empirical mass formulas.  For fermions,
spin 1/2 was assumed and for bosons spin 0 was assumed.  For each $Z$ we 
take $N=Z-1, Z,$ and $Z+1$.  We present in the Table I. average yields of
protons, neutrons, tritons, $^3$He, $^4$He and the sum of yields of
all nuclei with charges greater than 12.  Temperature range of 3 to 
6 MeV are of interest to many experiments.  We also show the results
at 30 MeV.  The CM approximation for composites
is seen to be quite good.

Granting that below a certain temperature, predictions from a grand canonical
model gets to be quite erroneous, could one predict when it becomes 
bad and why?  The answer to the first part is simple.  Usually,
the yield $<n_z>$ (or $<n_a>$ where $a$ is the mass number) falls
with $z$ but below a certain value of the temperature, 
the yield, after falling initially,
begins to rise again.  If this happens, one must discard the GCM
and do a CM.  The rise of yields, after reaching a minimum, signifies
several interesting features in intermediate energy multifragmentation
models.  In percolation and Lattice Gas model \cite{Bauer,Pan} this 
signifies the appearence of a percolating cluster.  In thermodynamic model,
the temperature at which this rise, after the minimum, just disappears
signifies a first order phase transition (in the infinite system and
no coulomb limit)
\cite{Dasgupta2}.  It suggests that at this temperature a large
blob of the system, usually identified as a liquid, has just disappeared.
It has been shown that at the transition temperature,
the specific heat at constant volume is very different in GCM and CM
although they match admirably at higher temperature
\cite{Das}.  The reason for
the discrepancy is an unusually large fluctuation in the number of
particles in the GCM below the transition temperature.  We have
however no fundamental understanding why such large fluctuations
appear in the GCM below the transition temperature.

This work is supported in part by the Natural Sciences and Engineering
Research Council of Canada.

\begin{table}
\begin{center}
\caption{ \it Comparision of claculations of average yields and E/A. By exact
we mean a calculation with proper symmetry.  Sum over discrete orbitals
in a box has been replaced by integration as is the usual practice.}
%\vspace {0.5in}
\begin{tabular}{cccccccccc}
\hline
\multicolumn{1}{c}{Calc} &
\multicolumn{1}{c}{$p$} &
\multicolumn{1}{c}{$n$} &
\multicolumn{1}{c}{$d$} &
\multicolumn{1}{c}{$t$} &
\multicolumn{1}{c}{$^{3}He$} &
\multicolumn{1}{c}{$^{4}He$} &
\multicolumn{1}{c}{$Z>12$} &
\multicolumn{1}{c}{Temp.} &
\multicolumn{1}{c}{$E/A$} \\
\hline
approx&0.307&0.032&0.050&0.007&0.054&0.679&0.945&3 $MeV$&-7.863 $MeV$ \\
exact&0.306&0.031&0.051&0.007&0.053&0.696&0.945&3 $MeV$&-7.861 $MeV$ \\
approx&1.174&0.898&1.177&0.560&0.641&2.489&0.051&6 $MeV$&-4.117 $MeV$ \\
exact&1.117&0.856&1.195&0.553&0.638&2.573&0.050&6 $MeV$&-4.135 $MeV$ \\
approx&4.127&3.955&4.812&2.099&2.052&1.985&0.000&12 $MeV$&4.401 $MeV$ \\
exact&3.860&3.696&4.941&2.090&2.051&2.021&0.000&12 $MeV$&4.308 $MeV$ \\
approx&10.937&10.893&7.664&1.686&1.650&0.379&0.000&30 $MeV$&28.914 $MeV$ \\
exact&10.512&10.468&7.885&1.732&1.696&0.395&0.000&30 $MeV$&28.844 $MeV$ \\
\hline
\end{tabular}
\end{center}
\end{table}

\end{document}